\documentclass{article}

\usepackage{spconf,graphicx}
\usepackage{bm,amsmath,amssymb,mathrsfs,amsfonts}
\usepackage{caption}
\usepackage{scalefnt}
\usepackage{multirow}
\usepackage{here}
\usepackage{booktabs}
\usepackage{enumitem}
\usepackage{type1cm}
\usepackage{cite}
\usepackage{url}
\usepackage{algorithm}
\usepackage{algpseudocode}
\usepackage{arydshln}
\usepackage{color}
\usepackage{graphics}
\usepackage{tikz}
\usepackage{pgfplots}
\pgfplotsset{compat=1.17}

\newcommand{\ie}{\textit{i.e.}}
\newcommand{\eg}{\textit{e.g.}}
\newcommand{\etal}{\textit{et~al.}}

\newcommand{\dmodel}{d_{\rm model}}
\newcommand{\dff}{d_{\rm ff}}
\newcommand{\nhead}{H}

\newcommand{\speech}{X}
\newcommand{\beamasr}{B_{\textsc{asr}}}
\newcommand{\beamst}{B_{\textsc{st}}}
\newcommand{\maskctciter}{K_{\rm mask}}
\newcommand{\certhres}{\theta_{\rm cer}}
\newcommand{\masktoken}{\texttt{[MASK]}}
\newcommand{\encasr}{{\rm Enc}_{\textsc{asr}}}
\newcommand{\decasr}{{\rm Dec}_{\textsc{asr}}}
\newcommand{\ctcasr}{{\rm CTC}_{\textsc{asr}}}
\newcommand{\encst}{{\rm Enc}_{\textsc{st}}}
\newcommand{\decst}{{\rm Dec}_{\textsc{st}}}
\newcommand{\hiddenasrenc}{h_{\textsc{asr}}}
\newcommand{\hiddenasrencinter}{h_{\textsc{asr}}^{l}}
\newcommand{\hiddenasrdec}{s_{\textsc{asr}}}
\newcommand{\hiddenasrdechat}{\hat{s}_{\textsc{asr}}}
\newcommand{\hiddenstenc}{h_{\textsc{st}}}
\newcommand{\hiddenstdec}{s_{\textsc{st}}}
\newcommand{\probctc}{P_{\textsc{ctc}}}
\newcommand{\probthres}{P_{\rm thres}}

\newcommand{\ysrc}{Y_{\rm src}}
\newcommand{\ytgt}{Y_{\rm tgt}}
\newcommand{\yhatctc}{\hat{Y}_{\textsc{ctc}}}
\newcommand{\yhatmask}{\hat{Y}_{\rm mask}}

\newcommand{\maskpredict}{{\rm MaskCTC}_{\textsc{asr}}}
\newcommand{\ctcgreedy}{{\rm Greedy}_{\textsc{ctc}}}
\newcommand{\beamsearchasr}{{\rm BeamSearch}_{\textsc{asr}}}

\newcommand{\lambdactc}{\lambda_{\textsc{ctc}}}
\newcommand{\lambdaasr}{\lambda_{\textsc{asr}}}
\newcommand{\lambdainter}{\lambda_{\rm Inter}}
\newcommand{\losstotal}{{\cal L}_{\rm total}}
\newcommand{\lossst}{{\cal L}_{\textsc{st}}}
\newcommand{\lossctc}{{\cal L}^{\textsc{ctc}}_{\textsc{asr}}}
\newcommand{\lossasr}{{\cal L}^{\textsc{trf}}_{\textsc{asr}}}
\newcommand{\lossinter}{{\cal L}_{\rm Inter}}
\newcommand{\positioninter}{S_{\rm Inter}}
\newcommand{\specaugfreq}{F_{\rm sp}}
\newcommand{\specaugtime}{T_{\rm sp}}
\newcommand{\specaugmasknumfreq}{M^{F}_{\rm sp}}
\newcommand{\specaugmasknumtime}{M^{T}_{\rm sp}}

\makeatletter
\def\adl@drawiv#1#2#3{%
        \hskip.5\tabcolsep
        \xleaders#3{#2.5\@tempdimb #1{1}#2.5\@tempdimb}%
                #2\z@ plus1fil minus1fil\relax
        \hskip.5\tabcolsep}
\newcommand{\cdashlinelr}[1]{%
  \noalign{\vskip\aboverulesep
           \global\let\@dashdrawstore\adl@draw
           \global\let\adl@draw\adl@drawiv}
  \cdashline{#1}
  \noalign{\global\let\adl@draw\@dashdrawstore
           \vskip\belowrulesep}}
\makeatother

\title{Fast-MD: Fast Multi-Decoder End-to-End Speech Translation\\with Non-Autoregressive Hidden Intermediates}
\name{Hirofumi Inaguma$^1$, Siddharth Dalmia$^2$, Brian Yan$^2$, Shinji Watanabe$^{2}$}
\address{$^1$Kyoto University, Japan $^2$Carnegie Mellon University, USA}
\begin{document}
\ninept
\maketitle
\begin{abstract}
The multi-decoder (MD) end-to-end speech translation model has demonstrated high translation quality by searching for better intermediate automatic speech recognition (ASR) decoder states as hidden intermediates (HI).
It is a two-pass decoding model decomposing the overall task into ASR and machine translation sub-tasks.
However, the decoding speed is not fast enough for real-world applications because it conducts beam search for both sub-tasks during inference.
We propose \textit{Fast-MD}, a fast MD model that generates HI by non-autoregressive (NAR) decoding based on connectionist temporal classification (CTC) outputs followed by an ASR decoder.
We investigated two types of NAR HI: (1) \textit{parallel HI} by using an autoregressive Transformer ASR decoder and (2) \textit{masked HI} by using Mask-CTC, which combines CTC and the conditional masked language model.
To reduce a mismatch in the ASR decoder between teacher-forcing during training and conditioning on CTC outputs during testing, we also propose sampling CTC outputs during training.
Experimental evaluations on three corpora show that Fast-MD achieved about 2$\times$ and 4$\times$ faster decoding speed than that of the naïve MD model on GPU and CPU with comparable translation quality.
Adopting the Conformer encoder and intermediate CTC loss further boosts its quality without sacrificing decoding speed.
\end{abstract}

\section{Introduction}\label{sec:intro}
End-to-end speech translation (E2E-ST) from source audio to target language text is an active research area for replacing conventional cascade systems~\cite{ney1999speech}, which stack automatic speech recognition (ASR) and machine translation (MT) models in a pipeline.
The E2E approach has several advantages over the cascade approach, such as robustness against ASR errors and low-latency inference because source transcription is bypassed.
However, E2E-ST models are generally affected by optimization because of a lack of training data and complex tasks, and many solutions have been investigated, such as pre-training and multi-task learning with auxiliary sub-tasks~\cite{weiss2017sequence,berard2018end,bansal-etal-2019-pre,bahar2019comparative}.

Beyond the simple multi-task learning sharing some of the parameters among the sub-tasks, two-pass decoding approaches with an intermediate ASR decoder have been studied~\cite{liu2020synchronous,le-etal-2020-dual,dong2021consecutive,anastasopoulos-chiang-2018-tied,sperber-etal-2019-attention,sung2019towards,dalmia-etal-2021-searchable}.
They exploit the compositionality~\cite{dalmia2019enforcing} in sequence tasks and take advantage of both E2E and cascade approaches.
They can also display transcripts and translations to users simultaneously.
These approaches can be categorized into interactive decoding~\cite{liu2020synchronous,le-etal-2020-dual,dong2021consecutive} and hierarchical decoding~\cite{anastasopoulos-chiang-2018-tied,sperber-etal-2019-attention,sung2019towards,dalmia-etal-2021-searchable}.
The multi-decoder (MD) architecture~\cite{dalmia-etal-2021-searchable,shi-etal-2021-highland} is attractive because it avoids intermediate ASR error propagation by dual cross-attention to both acoustic and textual encoder representations.
During inference, it can further fuse scores from an external language model (LM) and auxiliary connectionist temporal classification (CTC)~\cite{ctc_graves} module.
Despite the improved translation quality, however, the decoding speed lags behind that of vanilla E2E-ST models.
This is because of the introduction of an additional search space in the ASR sub-task, which is an additional fixed overhead.

Studies on fast decoding for the E2E-ST task have also attracted attention.
Most studies focused on the encoder and proposed to shrink encoder output lengths by using CTC probabilities~\cite{liu2020bridging,gaido-etal-2021-ctc,dong2021consecutive,zeng-etal-2021-realtrans} or $L_{0}$ regularization~\cite{zhang-etal-2020-adaptive}.
This is also important for bridging the modality gap between speech and text to make the most of pre-trained ASR and MT models~\cite{liu2020bridging,xu-etal-2021-stacked}.
A few studies instead adopted non-autoregressive (NAR) decoding at the cost of accuracy~\cite{inaguma2021orthros,chuang-etal-2021-investigating}.

Our goal in this study was to increase the decoding speed of the intermediate ASR module in the MD model while maintaining the benefits of better accuracy and compositionality.
We propose \textit{Fast-MD}, a fast MD model that performs NAR decoding led by CTC predictions in the ASR sub-net.
Instead of feeding the CTC output to the MT sub-net directly~\cite{liu2020bridging,xu-etal-2021-stacked}, we feed intermediate ASR decoder states as hidden intermediates (HI) after conditioning an additional Transformer ASR decoder on the CTC output in parallel (\textit{parallel HI}).
By doing this, we can avoid information loss while compressing the sequence length of acoustic representations.
If the decoding speed is not a concern, we can search for better HI by conducting beam search even with an external LM while maintaining the searchability of the MD architecture.
We also investigate the use of Mask-CTC~\cite{higuchi2020mask}, an iterative refinement-based NAR model with CTC and the conditional masked language model (CMLM)~\cite{ghazvininejad-etal-2019-mask}, as the ASR sub-net (\textit{masked HI}).

To reduce the gap in training and test conditions, where teacher-forcing~\cite{williams1989learning} is performed during training while CTC outputs are used to condition the ASR decoder at test time, we also introduce sampling CTC outputs from the CTC module during training, referred to as \textit{CTC sampling}.
We also improved the performance of CTC in Fast-MD by adopting the Conformer encoder~\cite{gulati2020} and intermediate CTC loss~\cite{lee2021intermediate}, which eventually improved the overall translation quality without sacrificing decoding speed.

Experimental evaluations on three benchmark datasets demonstrate that Fast-MD achieved about 2$\times$ and 4$\times$ faster decoding speed than that of the naïve MD model on a GPU and CPU without sacrificing translation quality.
The Conformer encoder and InterCTC loss also improved both recognition and translation accuracy.
We finally show that Fast-MD does not lose the searchability of better HI and is robust against long-form speech.

\vspace{-3mm}
\section{Background}

\vspace{-2mm}
\subsection{MD E2E-ST architecture}
\vspace{-1mm}
The MD architecture decomposes an E2E-ST model into ASR and MT sub-nets while retaining E2E differentiability.
Let $\speech$ be an input source speech sequence, $\ysrc$ be the ground-truth source transcript, and $\ytgt$ be the corresponding ground-truth target translation.
The ASR sub-net has the same structure as the Transformer ASR model with an auxiliary CTC layer $\ctcasr$~\cite{karita2019comparative} and generates continuous ASR decoder states $\hiddenasrdec$ as HI as follows:
\begin{eqnarray*}
\hiddenasrenc &=& \encasr(\speech), \\
\hiddenasrdec &=& \decasr(\hiddenasrenc),
\end{eqnarray*}
where $\encasr$, $\decasr$, and $\hiddenasrenc$ are an ASR encoder, ASR decoder, and encoder outputs, respectively.
The MT sub-net consists of a shallow ST encoder and ST decoder.
The ST encoder directly takes $\hiddenasrdec$ as inputs and generates textual representations $\hiddenstenc$.
Each block in the ST decoder has an additional cross-attention layer over $\hiddenasrenc$ (speech attention) right before a cross-attention layer over $\hiddenstenc$ as follows:
\begin{eqnarray*}
\hiddenstenc &=& \encst(\hiddenasrdec), \\
\hiddenstdec &=& \decst(\hiddenasrenc, \hiddenstenc),
\end{eqnarray*}
where $\encst$, $\decst$, and $\hiddenstdec$ are an ST encoder, ST decoder, and decoder states, respectively.
The speech attention is important for alleviating error propagation from the ASR sub-net.

The training objective is formulated as a linear interpolation of an E2E-ST cross-entropy (CE) loss $\lossst$, ASR CTC loss $\lossctc$, and ASR CE loss $\lossasr$ as follows:
\begin{multline}
\losstotal = (1- \lambda_{\textsc{asr}}) \lossst(\ytgt|\speech) \\
+ \lambda_{\textsc{asr}} ((1 - \lambdactc) \lossasr(\ysrc|\speech) + \lambdactc \lossctc(\ysrc|\speech)), \label{eq:total_loss_md}
\end{multline}
where $\lambda_{\rm *}$ is a loss weight.

During inference, the ASR sub-net performs beam search $\beamsearchasr$ with a beam width of $\beamasr$ to find better $\hiddenasrdec$.
The states corresponding to the most plausible transcript are fed to the MT sub-net.
If an external LM is available, we can also integrate the LM scores, which is effective for domain adaptation~\cite{dalmia-etal-2021-searchable}.
The complete decoding algorithm is shown in Algorithm~\ref{algo:md}.
To differentiate the inference from the Fast-MD inference, we refer to the former as \textit{Slow-MD inference}.

The introduction of the intermediate Transformer ASR decoder can compress \textit{frame-level} ASR encoder representations $\hiddenasrenc$ to \textit{token-level} HI $\hiddenasrdec$ ($|\hiddenasrdec| \ll |\hiddenasrenc|$) through multiple cross-attention operations.
This is more memory-efficient than feeding frame-level acoustic representations to the MT sub-net directly~\cite{xu-etal-2021-stacked} and enables injection of LM scores for searching better HI.
Compared with CTC-based shrinking~\cite{liu2020bridging,zeng-etal-2021-realtrans} and adaptive feature selection~\cite{zhang-etal-2020-adaptive}, information loss can be avoided thanks to the fine-grained cross-attention in $\decasr$ and an additional cross-attention in $\decst$.

\vspace{-2mm}
\subsection{NAR ASR}
\subsubsection{Mask-CTC}
Mask-CTC is an iterative NAR ASR model that combines CTC and the CMLM.
During inference, it first obtains CTC outputs $\yhatctc$ by greedy search $\ctcgreedy$, where the most plausible label is taken from the CTC layer at each time frame, and all blank labels are removed after collapsing repeated non-blank labels.
Next, less-confident tokens, the probabilities $\probctc$ of which are smaller than $\probthres$, are replaced with a mask token {\masktoken}, generating $\yhatmask$.
Then, the bidirectional Transformer decoder predicts tokens at the masked positions in $\yhatmask$ by leveraging surrounding contexts through a constant number of iterations $\maskctciter$ ($\geq 1$) as
\begin{eqnarray*}
\yhatctc &=& \ctcgreedy(\probctc(\speech)), \\
\yhatmask &=& \{y_{l} \in \yhatctc|\probctc(y_{l}|\speech) < \probthres\}, \\
\hiddenasrdechat &=& \maskpredict(\decasr(\hiddenasrenc), \yhatctc, \yhatmask, \maskctciter),
\end{eqnarray*}
where $\maskpredict$ is a search operation of Mask-CTC.

\vspace{-1mm}
\subsubsection{Intermediate CTC loss}
Intermediate CTC (InterCTC) loss $\lossinter$~\cite{lee2021intermediate} is a regularization method for CTC by calculating CTC losses at multiple intermediate encoder layers $\positioninter$ as
\begin{multline}
\lossinter = (1- \lambdainter) \lossctc(\ysrc|\hiddenasrenc) \\
+ \lambdainter (\frac{1}{|\positioninter|} \sum_{l \in \positioninter} \lossctc(\ysrc|\hiddenasrencinter)),  \label{eq:inter_ctc}
\end{multline}
where $\hiddenasrencinter$ are the encoder outputs at the $l$-th layer, and $\lambdainter$ is a weight for the InterCTC loss.
During inference, only the CTC probabilities from the top layer are used; therefore, InterCTC does not slow the decoding speed.

\begin{algorithm}[t]
\caption{Decoding algorithm of MD and Fast-MD}
\footnotesize
\begin{algorithmic}[1]
\Function{Decode}{$\speech$, $\beamasr$, $\beamst$, $\maskctciter$, $\probthres$}
    \State $\hiddenasrenc \gets \encasr(X)$
    \State $\probctc \gets \ctcasr(\hiddenasrenc)$
    \State

    \If{\textit{$fastmd\_parallel$}}
        \State {\color{blue} // Fast-MD (parallel HI), $O(1)$} \label{algo:fastmd_parallel_start}
        \State $\yhatctc \gets \ctcgreedy(X)$  \label{algo:ctc_greedy_decoding}
        \State $\hiddenasrdechat \gets {\color{red} {\rm TeacherForce}_{\textsc{asr}}}(\decasr(\hiddenasrenc), \yhatctc)$ \label{algo:get_nar_hidden_parallel} \label{algo:fastmd_parallel_end}

    \ElsIf{\textit{$fastmd\_masked$}}
        \State {\color{blue} // Fast-MD (masked HI), $\maskctciter \times O(1)$} \label{algo:fastmd_masked_start}
        \State $\yhatctc \gets \ctcgreedy(X)$
        \State $\yhatmask \gets \{y_{l} \in \yhatctc| \probctc(y_{l}| \speech) < \probthres\}$
        \State $\hiddenasrdechat \gets {\color{red} \maskpredict}(\decasr(\hiddenasrenc), \yhatctc, \yhatmask, \maskctciter)$ \label{algo:get_nar_hidden_maskctc} \label{algo:fastmd_masked_end}
    \Else
        \State {\color{blue} // Slow-MD inference, $O(N)$ or $O(NT)$}  \label{algo:md_start}
        \State $\hiddenasrdechat \gets {\color{red} \beamsearchasr}(\decasr(\hiddenasrenc), \probctc, \beamasr)$  \label{algo:md_end}
    \EndIf
    \State

    \State $\hiddenstenc \gets \encst(\hiddenasrdechat)$
    \State \Return ${\rm BeamSearch}_{\textsc{st}}(\decst(\hiddenasrenc, \hiddenstenc), \beamst)$ {\color{blue} \Comment{$O(N)$}}
\EndFunction
\end{algorithmic}\label{algo:md}
\end{algorithm}
\setlength{\textfloatsep}{4.0mm}  %

\vspace{-2mm}
\section{Fast-MD}
In this section, we explain Fast-MD for increasing the decoding speed of the MD model by generating intermediate ASR decoder states, HI, in a non-autoregressive manner.

\vspace{-2mm}
\subsection{NAR HI}
Although the MD architecture improves translation quality, the decoding speed is much slower than a vanilla E2E-ST model because beam search is conducted in both source and target language spaces.
The optimal beam width of the ASR sub-net $\beamasr$ is relatively larger than that of the MT sub-net $\beamst$ ({\eg}, $\beamasr=16$ vs. $\beamst=10$ in~\cite{dalmia-etal-2021-searchable}), and integration of CTC and LM scores also makes the ASR decoding more complex~\cite{hybrid_ctc_attention}.
Therefore, search in the ASR sub-net is a dominant factor for the overall decoding speed in the naïve MD model.
Motivated by this, we focus on improving the decoding speed of the ASR sub-net while minimizing translation-quality degradation.
To this end, we have \textit{parallel HI} and \textit{masked HI} versions of Fast-MD by extracting HI by NAR decoding led by CTC.

\vspace{-2mm}
\subsubsection{Parallel HI}
The first version of Fast-MD obtains HI in parallel from an AR Transformer ASR decoder conditioned on CTC outputs.
This does not change the MD architecture, but HI are obtained in a different manner at test time.
Unlike the naïve MD model, Fast-MD conducts greedy search with $\ctcasr$ instead of beam search with $\decasr$ (\textit{line:\ref{algo:ctc_greedy_decoding}} in Algorithm~\ref{algo:md}).
The resulting CTC outputs $\yhatctc$ are used to obtain HI by conditioning $\decasr$ on them, {\ie}, teacher-forcing (\textit{line:\ref{algo:get_nar_hidden_parallel}}), as
\begin{eqnarray}
\hiddenasrdechat \gets {\rm TeacherForce}_{\textsc{asr}}(\decasr(\hiddenasrenc), \yhatctc), \label{eq:parallel_hi}
\end{eqnarray}
where ${\rm TeacherForce}_{\textsc{asr}}$ is a teacher-forcing operation.
We refer to the HI obtained with Eq.~\eqref{eq:parallel_hi} as \textit{parallel HI}.
Although $\decasr$ is still an AR model, we regard parallel HI as a family of NAR HI because we can feed $\yhatctc$ to $\decasr$ in parallel at all positions~\cite{song2021non,fan2021cass,tian2020one,wang2021wnars,zhang2020unified}.

\vspace{-2mm}
\subsubsection{Masked HI}
The second version of Fast-MD uses Mask-CTC as $\decasr$ instead of the AR Transformer decoder as follows:
\begin{eqnarray}
\hiddenasrdechat \gets \maskpredict(\decasr(\hiddenasrenc), \yhatctc, \yhatmask, \maskctciter). \label{eq:masked_hi}
\end{eqnarray}
We refer to the HI obtained from the Mask-CTC decoder as \textit{masked HI}.
In this case, we can improve $\yhatctc$ through $\maskctciter$ iterations, which corresponds to \textit{line~\ref{algo:fastmd_masked_start}-\ref{algo:fastmd_masked_end}} in Algorithm~\ref{algo:md}.
We use HI right after the last iteration.
Therefore, some decoder input tokens used to obtain the HI would be {\masktoken} in accordance with the CTC confidence scores.
This is desirable to avoid error propagation from CTC because such low-confident predictions are more likely to be wrong.
Therefore, the ST decoder should rely on $\hiddenasrenc$ more to complement the masked parts.

\vspace{-2mm}
\subsection{ASR error robust training}
While NAR ASR decoding can accelerate decoding speed, recognition accuracy still lags behind that of the AR decoding~\cite{higuchi2020mask,higuchi2021improved}.
Moreover, the ST encoder in the naïve MD model always consumes HI obtained by teacher-forcing with ground-truth transcripts during training.
Therefore, the NAR decoding in the ASR sub-net would enlarge the gap in training and test conditions and ultimately degrade the subsequent translation quality.

\vspace{-2mm}
\subsubsection{CTC sampling}
To overcome this problem, we introduce \textit{CTC sampling}, where greedy CTC outputs $\yhatctc$ are sampled on the fly during training and used to condition $\decasr$.\footnote{Concurrently, CTC sampling was also investigated in~\cite{song2021non} to improve the recognition performance of NAR ASR models. However, our motivation to introduce CTC sampling is to make the MD architecture robust against errors from the intermediate ASR component.}
We carry out greedy search with $\ctcasr$ for each utterance and use the resulting tokens $\yhatctc$ to obtain HI $\hiddenasrdechat$.
These HI are fed to $\encst$, and $\lossst$ is calculated accordingly.
For masked HI, we first sample $\yhatctc$ then apply a random mask to them.
This was more effective than generating a token mask based on CTC confidence scores in our preliminary experiments.
Note that we must pass source language tokens to the ASR decoder twice to calculate $\lossasr$ and obtain $\hiddenasrdechat$.
However, because the greedy search with CTC is fast enough and we can feed $\yhatctc$ to $\decasr$ in parallel, the additional computation time for CTC sampling is small (+38\%).
Although the argmax operation in CTC sampling is not differentiable, we can still keep a gradient path from $\encst$ to $\encasr$ via cross-attention in $\decasr$.

\subsubsection{Character-error-rate thresholding}
In the early training stage, the greedy CTC outputs $\yhatctc$ would not be very accurate.
Therefore, it would have a negative impact on the training, especially when training all parameters from random initialization.
To control the quality of $\yhatctc$, we filter out noisy $\yhatctc$ by thresholding on the basis of the character error rate (CER).
Specifically, for each utterance in every mini-batch, we calculate the CER of $\yhatctc$ with the corresponding $\ysrc$.
If the CER is larger than $\certhres$, we replace $\yhatctc$ in Eq.~\eqref{eq:parallel_hi} and Eq.~\eqref{eq:masked_hi} with $\ysrc$.

\vspace{-2mm}
\subsection{Improvements in CTC}
Because the final translation quality of Fast-MD is dependent of CTC accuracy, it is reasonable to expect that the better the CTC performance, the better the translation quality we can obtain.
Dejavu~\cite{tjandra2020deja}, InterCTC, and self-conditioned CTC~\cite{nozaki2021relaxing} improve CTC performance by enhancing intermediate encoder representations by calculating CTC losses for intermediate layers as well.
Adapting a better encoder architecture, such as Conformer~\cite{gulati2020}, is also simple yet effective~\cite{higuchi2021improved,ng2021pushing}.
We investigated combining Fast-MD with the Conformer encoder and InterCTC loss and studied how important they are to improve translation quality.

\begin{table*}[t]
    \centering
    \begingroup
    \caption{BLEU scores on \underline{Fisher-CallHome Spanish Es$\to$En}. Decoding speed was measured with batch size of 1 on Fisher-\texttt{test} set.}
    \label{tab:asru2021_result_fisher}
    \vspace{-3mm}
    \scalebox{0.92}{
    \begin{tabular}{lcccccccccc} \toprule
      \multirow{3}{*}{Model} & \multirow{4}{*}{\shortstack{Intermediate\\decoding\\complexity}} & \multicolumn{5}{c}{BLEU ($\uparrow$)} & \multicolumn{2}{c}{Decoding speed} \\ \cmidrule(lr){3-7} \cmidrule(lr){8-9}
       & & \multicolumn{3}{c}{Fisher} & \multicolumn{2}{c}{CallHome} & \multicolumn{2}{c}{GPU / CPU} \\ \cmidrule(lr){3-5} \cmidrule(lr){6-7}
       & & dev & dev2 & test & devtest & evltest & RTF ($\downarrow$) & Speedup ($\uparrow$) \\
      \midrule
        \textbf{Previous studies} \\
            \ ESPnet-ST~\cite{inaguma-etal-2020-espnet} & -- & 48.9 & 49.3 & 48.4 & 18.8 & 18.7 & -- & -- \\
            \ Cascade~\cite{dalmia-etal-2021-searchable} & -- & 50.4 & 51.2 & 50.7 & 19.6 & 19.2 & -- & -- \\

            \ Transformer MD~\cite{dalmia-etal-2021-searchable} ($\beamst$=10) & $O(N)$ & 54.6 & 54.6 & 54.1 & 21.7 & 21.4 & 0.12 / 4.84 & 1.00$\times$ / 1.00$\times$ \\
            \ \ + CTC scores + LM fusion & $O(NT)$ & 55.2 & 55.2 & 55.0 & 21.7 & 21.5 & -- & -- \\  %
            \midrule[\heavyrulewidth]

            \textbf{Transformer encoder} \\
            \ Vanilla E2E ($\beamst$=4) & -- & 49.4 & 50.6 & 49.4 & 18.6 & 18.4 & 0.05 / 1.06 & 2.59$\times$ / 4.55$\times$ \\
            \ MD ($\beamasr$=16, $\beamst$=4) & $O(N)$ & 53.8 & 54.5 & 54.3 & 21.5 & 21.5 & 0.10 / 3.52 & 1.16$\times$ / 1.38$\times$ \\
             \ \ + $\beamasr$=1 & $O(N)$ & 53.1 & 53.7 & 53.2 & 21.1 & 20.8 & 0.08 / 1.46 & 1.52$\times$ / 3.32$\times$ \\
             \ \ + Fast-MD inference & $O(1)$ & 51.8 & 52.4 & 52.3 & 19.7 & 19.7 & 0.06 / 1.20 & 2.00$\times$ / 4.02$\times$ \\

            \ Fast-MD (parallel HI) & $O(1)$ & \bf{54.0} & \bf{54.8} & \bf{53.9} & \bf{20.9} & \bf{20.7} & {\bf 0.06} / {\bf 1.20} & {\bf 2.00}$\times$ / {\bf 4.02}$\times$ \\ %
            \ Fast-MD (masked HI) & $O(1)$ & \bf{54.8} & \bf{55.1} & \bf{54.4} & {\bf 21.3} & \bf{21.3} & {\bf 0.06} / {\bf 1.24} & {\bf 1.94}$\times$ / {\bf 3.92}$\times$ \\ %
            \bottomrule
    \end{tabular}
    }
    \vspace{-3mm}
    \endgroup
\end{table*}

\vspace{-1mm}
\section{Experimental evaluation}
\subsection{Experimental setting}
\subsubsection{Dataset}
\noindent\textbf{Fisher-CallHome Spanish}~\cite{fisher_callhome} (Es$\to$En) contains 170 h of Spanish conversational telephone speech as well as the transcripts and English translations.
All punctuation marks except for apostrophes were removed from both transcripts and translations following the standard practice of this corpus~\cite{weiss2017sequence}.
We report case-insensitive detokenized BLEU scores~\cite{papineni-etal-2002-bleu} on Fisher-\{\texttt{dev}, \texttt{dev2}, \texttt{test}\}, and CallHome-\{\texttt{devtest}, \texttt{evltest}\}.
Note that the Fisher splits were evaluated with four English references.

\vspace{2mm}
\noindent\textbf{Libri-trans}~\cite{libri_trans} (En$\to$Fr) includes 100 h of English read speech as well as the transcripts and French translations.
Each translation in the official training set is augmented with Google Translate, and we used two French translations per speech utterance.
We report case-insensitive detokenized BLEU scores on the \texttt{test} set.

\vspace{2mm}
\noindent\textbf{Must-C}~\cite{mustc} contains English speech extracted from TED talks as well as the transcripts and translations.
We used the En$\to$De part of this corpus, which has 408 h of speech data.
Non-verbal speech labels such as ``\textit{(Applause)}'' and  ``\textit{(Laughter)}'' were removed during evaluation.
We report case-sensitive detokenized BLEU scores on the \texttt{tst-COMMON} set.

\vspace{-2mm}
\subsubsection{Pre-processing}
We conducted experiments on the basis of recipes in the ESPnet-ST toolkit~\cite{inaguma-etal-2020-espnet}.
We extracted 80-dimensional log-mel filterbank coefficients computed with a 25-ms window and shifted every 10-ms with three-dimensional pitch features with the Kaldi toolkit~\cite{kaldi}.
We augmented speech data with three-fold speed perturbation~\cite{speed_perturbation} and SpecAugment~\cite{specaugment}.
We used SpecAugment hyperparameters of $(\specaugmasknumtime, \specaugmasknumfreq, \specaugtime, \specaugfreq)=(2, 2, 40, 30)$ but did not use time-warping.
We removed utterances having more than 3000 speech frames or more than 400 characters to fit the GPU memory.

All sentences (both transcripts and translations) were tokenized with the {\tt tokenizer.perl} script in the Moses toolkit~\cite{koehn-etal-2007-moses}.
Punctuation marks except for apostrophes and case information were removed from all source transcripts.
We constructed joint source and target vocabularies on the basis of the byte pair encoding algorithm~\cite{sennrich-etal-2016-neural} with the Sentencepiece toolkit~\cite{kudo-richardson-2018-sentencepiece}.
However, we built the vocabulary on the basis of transcripts only for the ASR sub-net.
One thousand units were used for both ASR and MT sub-nets on Fisher and Libri-trans while 5k and 8k units were used for ASR and MT sub-nets on Must-C, respectively.

\vspace{-2mm}
\subsubsection{Architecture}
\vspace{-1mm}
We used Transformer~\cite{vaswani2017attention} and Conformer~\cite{gulati2020} models, and the implementations were based on previous studies~\cite{karita2019comparative,guo2021recent}.
The only difference between the Transformer and Conformer models is the speech encoder, and the rest of the architecture is the same.
Following Dalmia~{\etal}'s study~\cite{dalmia-etal-2021-searchable}, we used 12 ASR encoder blocks, 6 ASR decoder blocks, 2 ST encoder blocks, and 6 ST decoder blocks.
We increased the number of ST encoder blocks to four on Must-C.
The ASR encoder had 2 CNN blocks with a kernel size of 3 and channel size of 256 before the first encoder block, resulting in four-fold down-sampling on both time and frequency axes.
The dimensions of the self-attention layer $\dmodel$ and feed-forward network $\dff$ were set to 256 and 2048 in all sub-nets, respectively.
The number of attention heads $\nhead$ was set to four.
For the Conformer encoder, we set the kernel size of depthwise separable convolution in each convolutional module to 15.

Unlike Dalmia~{\etal}’s study~\cite{dalmia-etal-2021-searchable}, where both ASR and MT sub-nets shared the same joint source and target vocabulary, we used the \textit{source-language-only} vocabulary for the ASR sub-net.
We found that this improves both ASR and ST performances.

\vspace{-2mm}
\subsubsection{Training}
\vspace{-1mm}
The Adam optimizer~\cite{adam} with $\beta_{1}=0.9$, $\beta_{2}=0.98$, and $\epsilon=10^{-9}$ was used for training with a Noam learning rate schedule~\cite{vaswani2017attention}.
The warmup step was set to 25k.
We trained all MD models for 50 epochs with an effective batch size of 384 utterances and learning-rate scale of 12.5.
Regularization hyperparameters, such as dropout rate and label smoothing probability, were the same as those in Dalmia~{\etal}'s study~\cite{dalmia-etal-2021-searchable}.
We set $\lambdaasr$ and $\lambdactc$ in Eq.~\eqref{eq:total_loss_md} to 0.5 and 0.3, respectively.
When using InterCTC loss, $\lossctc$ in Eq.~\eqref{eq:total_loss_md} was replaced with $\lossinter$ in Eq.~\eqref{eq:inter_ctc} with $\lambdainter=0.3$.
We tuned $\certhres$ from \{0.2, 0.3, 0.4\} on the basis of the validation BLEU score.
MD models were trained from random initialization~\cite{dalmia-etal-2021-searchable}, except for Fast-MD (masked HI), the ASR sub-net of which was initialized with a pre-trained Mask-CTC ASR model.
The last ten checkpoints were used for model averaging of the MD models.

\vspace{-2mm}
\subsubsection{Decoding}
\vspace{-1mm}
The Slow-MD inference used $\beamasr=16$ without CTC or LM scores unless otherwise specified.
For both Fast-MD and Slow-MD inferences, we used $\beamst=4$, where $\beamst$ was reduced from 10 to 4 unlike in Dalmia~{\etal}’s study~\cite{dalmia-etal-2021-searchable}.
The $\maskctciter$ was set to 1 for Mask-CTC in Fast-MD (masked HI).
Therefore, both Fast-MD (parallel HI) and Fast-MD (masked HI) have the same decoding complexity theoretically.
The decoding speed was measured with a batch size of 1 on an NVIDIA TITAN RTX GPU and Intel(R) Xeon(R) Gold 6128 CPU @ 3.4GHz by averaging on five runs.
Note that beam search was implemented efficiently by updating decoder states with a batch-level matrix multiplication~\cite{seki2019}.
We calculated detokenized BLEU scores with SacreBLEU~\cite{post-2018-call}.

\vspace{-1mm}
\subsection{Main results: Fisher-CallHome Spanish Es$\to$En}\label{ssec:main_results}
\subsubsection{ST performance evaluation}
The results of the ST performance are shown in Table~\ref{tab:asru2021_result_fisher}.
Note that we re-implemented the naïve MD model and were able to reproduce similar BLEU scores.
When reducing the search space of the ASR sub-net in the naïve MD model by greedy search ($\beamasr=1$) or the Fast-MD inference, we observed severe degradation in BLEU scores by about 2 BLEU.
This is because the model did not recognize ASR errors during training, which was our motivation to introduce CTC sampling.
Fast-MD (parallel HI), however, could recover the degradation by 80\% and achieved 1.72$\times$ and 2.91$\times$ faster decoding over the MD model ($\beamst=4$) on the GPU and CPU, respectively.
The speed increase was much larger on the CPU, which is important for on-device applications.

Fast-MD (masked HI) outperformed Fast-MD (parallel HI) in BLEU scores.
It even surpassed the naïve MD model on the Fisher test sets.
A possible explanation of this behavior is that the MT sub-net could ignore less-confident and noisy predictions from the ASR sub-net, which overcame ASR error propagation effectively.
Because of $\maskctciter=1$, the decoding cost is almost same as Fast-MD (parallel HI).
One limitation of masked HI is that any successful method to fuse external LM scores to Mask-CTC has not been established, which we leave future work.
Compared with the vanilla E2E model, Fast-MD achieved much higher BLEU scores with a small decrease of the decoding speed.

\begin{table}[t]
    \centering
    \caption{Improving CTC in MD models with Conformer encoder and intermediate CTC loss on Fisher-CallHome Spanish corpus}
    \label{tab:asru2021_improve_ctc}
    \vspace{-3mm}
    \begingroup
    \scalebox{0.86}{
    \begin{tabular}{lccccc} \toprule
    \multirow{3}{*}{Model} & \multicolumn{5}{c}{BLEU ($\uparrow$)} \\ \cmidrule{2-6}
    & \multicolumn{3}{c}{Fisher} & \multicolumn{2}{c}{CallHome} \\ \cmidrule(lr){2-4} \cmidrule(lr){5-6}
    & dev & dev2 & test & devtest & evltest \\
    \midrule
       \ MD & 53.8 & 54.5 & 54.3 & 21.5 & 21.5 \\
       \ \ + Conformer Enc. & 55.0 & 55.3 & 54.6 & 22.3 & 21.5 \\
       \ \ \ ++ InterCTC loss & \bf{55.0} & \bf{55.6} & \bf{55.6} & \bf{22.4} & \bf{22.1} \\
       \cmidrule(lr){1-6}

       \ MD + Fast-MD inference & 51.8 & 52.4 & 52.3 & 19.7 & 19.7 \\
       \ \ + Conformer Enc. & 55.2 & 55.6 & 54.8 & 22.1 & 21.6 \\
       \ \ \ ++ InterCTC loss & \bf{55.3} & \bf{55.7} & \bf{55.6} & \bf{22.6} & \bf{22.0} \\
       \cmidrule(lr){1-6}

       \ Fast-MD (parallel HI) & 54.0 & 54.8 & 53.9 & 20.9 & 20.7 \\  %
       \ \ + Conformer Enc. & 55.7 & 56.2 & 55.2 & 22.6 & \bf{22.3} \\  %
       \ \ \ ++ InterCTC loss & \bf{56.3} & \bf{56.4} & \bf{56.2} & \bf{23.3} & 22.2 \\  %
       \cmidrule(lr){1-6}

       \ Fast-MD (masked HI) & 54.8 & 55.1 & 54.4 & 21.3 & 21.3 \\ %
       \ \ + Conformer Enc. & 55.2 & 56.0 & 55.3 & 22.2 & 22.0 \\  %
       \ \ \ ++ InterCTC loss & \bf{56.3} & \bf{57.4} & \bf{56.2} & \bf{22.8} & \bf{22.8} \\  %
     \bottomrule
    \end{tabular}
    }
    \endgroup
    \vspace{-1mm}
\end{table}

\begin{table}[t]
    \centering
    \caption{Word error rate (WER) of ASR sub-net in naïve MD model on Fisher-CallHome Spanish corpus}
    \label{tab:asru2021_evaluate_asr_subnet}
    \vspace{-3mm}
    \begingroup
    \scalebox{0.86}{
    \begin{tabular}{lccccc} \toprule
    \multirow{3}{*}{Model} & \multicolumn{5}{c}{WER ($\downarrow$)} \\ \cmidrule{2-6}
    & \multicolumn{3}{c}{Fisher} & \multicolumn{2}{c}{CallHome} \\ \cmidrule(lr){2-4} \cmidrule(lr){5-6}
    & dev & dev2 & test & devtest & evltest \\
    \midrule
     \textbf{Slow-MD inference} \\
     \ Transformer MD & 23.1 & 22.3 & 20.4 & 38.8 & 39.5 \\
     \ \ + Conformer Enc. & 23.1 & 23.0 & 21.4 & 37.6 & 38.9 \\
     \ \ \ ++ InterCTC loss & \bf{22.2} & \bf{21.5} & \bf{20.1} & \bf{36.8} & \bf{38.2} \\
     \midrule

     \textbf{Fast-MD inference} \\
     \ Transformer MD & 26.2 & 25.6 & 23.5 & 43.1 & 43.6 \\
     \ \ + Conformer Enc. & 21.4 & 21.2 & 19.6 & 37.5 & 37.9 \\
     \ \ \ ++ InterCTC loss & \bf{20.7} & \bf{20.5} & \bf{19.1} & \bf{36.5} & \bf{36.2} \\
     \bottomrule
    \end{tabular}
    }
    \endgroup
\end{table}

\vspace{-1mm}
\subsubsection{Does better CTC lead to better ST performance?}
Next, we investigated the effectiveness of the Conformer encoder and InterCTC loss for MD and Fast-MD.
For Fast-MD (masked HI), we applied InterCTC loss to ASR pre-training as well.
The results in Table~\ref{tab:asru2021_improve_ctc} indicate that the Conformer encoder improved the BLEU scores in all the settings by a large margin.
Interestingly, the Fast-MD inference outperformed the Slow-MD inference for the Conformer-based MD.
This was because the CTC performance was better than that of the Transformer ASR decoder, as mentioned in Section~\ref{sssec:asr_evaluation}.
However, we confirmed that CTC sampling in the Conformer-based Fast-MD was still effective.
Parallel HI was as effective as masked HI when using the Conformer encoder, unlike the results shown in Table~\ref{tab:asru2021_result_fisher}.
To the best of our knowledge, the best Conformer-based Fast-MD models achieved state-of-the-art BLEU scores on all test sets.

\vspace{-1mm}
\subsubsection{ASR performance evaluation}\label{sssec:asr_evaluation}
We evaluate the word error rate (WER) of the ASR sub-net in the naïve MD models.
The results in Table~\ref{tab:asru2021_evaluate_asr_subnet} indicate that the Conformer encoder and InterCTC reduced WER consistently regardless of the search method.
However, CTC greedy search with the Conformer encoder outperformed beam search with the Transformer decoder, which was consistent with the observations in Table~\ref{tab:asru2021_improve_ctc}.
Therefore, we conclude that improvement in the ASR sub-net can improve overall translation quality.
We also observed a similar trend in WER for the Fast-MD models.

\begin{table}[t]
    \centering
    \begingroup
    \caption{BLEU scores on \underline{Libri-trans En$\to$Fr} \texttt{test} set. *Tokenized BLEU scores. For Slow-MD inference, CTC scores were also used.}
    \label{tab:asru2021_result_libritrans}
    \vspace{-3mm}
    \scalebox{0.86}{
    \begin{tabular}{lccc} \toprule
      \multirow{2}{*}{Model} & \multirow{2}{*}{BLEU ($\uparrow$)} & \multirow{2}{*}{WER ($\downarrow$) }& Speedup ($\uparrow$) \\
       & & & GPU / CPU \\
      \midrule
            \textbf{Previous studies} \\
            \ ESPnet-ST~\cite{inaguma-etal-2020-espnet} & 16.8${}^{\phantom{0}}$ & -- & -- \\
            \ WordKD~\cite{liu2019end} & 17.0* & -- & -- \\
            \ TCEN-LSTM~\cite{wang2020bridging} & 17.0${}^{\phantom{0}}$ & -- & -- \\
            \ NeurST~\cite{zhao-etal-2021-neurst} & 17.2${}^{\phantom{0}}$ & -- & -- \\
            \ Curriculum PT~\cite{wang-etal-2020-curriculum} & 17.7* & -- & -- \\
            \ LUT~\cite{dong2021listen} & 17.8* & -- & -- \\
            \ STAST~\cite{liu2020bridging} & 17.8* & -- & -- \\
            \ COSTT~\cite{dong2021consecutive} & 17.8${}^{\phantom{0}}$ & -- & -- \\
            \ SATE~\cite{xu-etal-2021-stacked} & 18.3${}^{\phantom{0}}$ & -- & -- \\
            \midrule[\heavyrulewidth]

            \textbf{Transformer encoder} \\
            \ MD & 18.2${}^{\phantom{0}}$ & \phantom{0}6.8 & 1.00$\times$ / 1.00$\times$ \\
            \ \ + Fast-MD inference & 17.4${}^{\phantom{0}}$ & 10.7 & 4.98$\times$ / 3.98$\times$ \\
            \ Fast-MD (parallel HI) & \bf{18.1}${}^{\phantom{0}}$ & 10.5 & {\bf 4.98}$\times$ / {\bf 3.98}$\times$ \\  %
            \ \ + Slow-MD inference & \bf{18.3}${}^{\phantom{0}}$ & \phantom{0}6.7 & 1.00$\times$ / 1.00$\times$ \\
            \ Fast-MD (masked HI) & \bf{18.5}${}^{\phantom{0}}$ & \phantom{0}9.0 & {\bf 4.88}$\times$ / {\bf 3.96}$\times$ \\
            \cmidrule{1-4}

            \textbf{Conformer encoder} \\
            \ MD & 18.5${}^{\phantom{0}}$ & \phantom{0}6.2 & 1.00$\times$ / 1.00$\times$ \\
            \ \ + Fast-MD inference & 18.3${}^{\phantom{0}}$ & \phantom{0}5.8 & 4.85$\times$ / 4.17$\times$ \\

            \ Fast-MD (parallel HI) & 18.7${}^{\phantom{0}}$ & \phantom{0}5.6 & {\bf 4.85}$\times$ / {\bf 4.17}$\times$ \\  %
            \ \ + InterCTC & \bf{18.8}${}^{\phantom{0}}$ & \phantom{0}\bf{5.2} & {\bf 4.85}$\times$ / {\bf 4.17}$\times$ \\  %
            \ Fast-MD (masked HI) & 18.6${}^{\phantom{0}}$ & \phantom{0}5.5 & {\bf 5.00}$\times$ / {\bf 4.20}$\times$ \\
            \ \ + InterCTC & \bf{19.1}${}^{\phantom{0}}$ & \phantom{0}\bf{5.0}& {\bf 5.00}$\times$ / {\bf 4.20}$\times$ \\  %
            \bottomrule
    \end{tabular}
    }
    \endgroup
\end{table}

\begin{table}[t]
    \centering
    \tabcolsep 4pt
    \begingroup
    \caption{BLEU scores on \underline{Must-C En$\to$De} \texttt{tst-COMMON} set}
    \label{tab:asru2021_result_mustc}
    \vspace{-3mm}
    \scalebox{0.86}{
    \begin{tabular}{lccc} \toprule
      \multirow{2}{*}{Model} & \multirow{2}{*}{BLEU ($\uparrow$)} & \multirow{2}{*}{WER ($\downarrow$)} & Speedup ($\uparrow$) \\
      & & & GPU / CPU \\
      \midrule
            \textbf{Previous studies} \\
            \ ESPnet-ST~\cite{inaguma-etal-2020-espnet} & 22.9 & -- & -- \\
            \ Fairseq S2T~\cite{wang-etal-2020-fairseq} & 22.7 & -- & -- \\
            \ NeurST~\cite{zhao-etal-2021-neurst} & 22.8 & -- & -- \\
            \ AFS${}^{t,f}$~\cite{zhang-etal-2020-adaptive} & 22.4 & -- & -- \\
            \ STAST~\cite{liu2020bridging} & 23.1 & -- & -- \\
            \ Deep Relative Transformer~\cite{pham2020relative} & 25.2 & -- & -- \\
            \ SATE~\cite{xu-etal-2021-stacked} & 25.2 & -- & -- \\
            \ Transformer MD~\cite{dalmia-etal-2021-searchable} & 26.4 & -- & -- \\
            \ Conformer + SeqKD (2ref)~\cite{inaguma-etal-2021-source} & 26.8 & -- & -- \\
            \ Conformer + Bidir SeqKD~\cite{inaguma-etal-2021-source} & 27.0 & -- & -- \\
            \midrule[\heavyrulewidth]

            \textbf{Conformer encoder} \\
            \ MD & 26.0 & 11.1 & 1.00$\times$ / 1.00$\times$ \\
            \ \ + Fast-MD inference & 25.9 & 11.2 & 1.70$\times$ / 2.47$\times$ \\
            \ Fast-MD (parallel HI) & 26.4 & 10.9 & {\bf 1.70}$\times$ / {\bf 2.47}$\times$ \\  %
            \ \ + InterCTC & \bf{26.5} & \bf{10.8} & {\bf 1.70}$\times$ / {\bf 2.47}$\times$ \\  %
            \bottomrule
    \end{tabular}
    }
    \vspace{-1mm}
    \endgroup
\end{table}

\vspace{-2mm}
\subsection{Results on Libri-trans: En$\to$Fr}\label{ssec:libritrans}
The results of Libri-trans are shown in Table~\ref{tab:asru2021_result_libritrans}.
In this corpus, we needed to use auxiliary CTC scores during inference for the ASR sub-net to achieve decent WER because of many long-form speech utterances in the test set.
Similar to the conversational domain, we confirmed that Fast-MD (parallel HI) improved by 0.7 BLEU and 0.4 BLEU compared with the naïve MD model with the mismatched Fast-MD inference for the Transformer and Conformer encoders, respectively.
Because this corpus has longer utterances than the Fisher-CallHome Spanish corpus, the relative increase in decoding speed was much larger.
Fast-MD (masked HI) further improved both recognition and translation performances with similar decoding speeds.
InterCTC loss advanced both Fast-MD (parallel HI) and Fast-MD (masked HI) further.
Regarding WER, the Slow-MD inference outperformed the Fast-MD inference when the Transformer encoder was used.
When using the Conformer encoder, CTC greedy search outperformed beam search with the Transformer ASR decoder by 0.4, which was consistent with the observations on Fisher-CallHome Spanish.
Interestingly, the WER of CTC greedy search also improved by using CTC sampling.

\vspace{-2mm}
\subsection{Results on Must-C: En$\to$De}\label{ssec:mustc}
We show the results on Must-C in Table~\ref{tab:asru2021_result_mustc} to confirm the effectiveness of Fast-MD in a large dataset.
We focused on Fast-MD (parallel HI) to avoid ASR pre-training.
In this corpus, BLEU scores of the naïve MD model did not degrade much with the mismatched Fast-MD inference, which can also be confirmed from the similar WER.
However, Fast-MD improved upon the naïve MD model by 0.4 BLEU with 1.70$\times$ and 2.47$\times$ faster decoding on the GPU and CPU, respectively.
Note that the MD/Fast-MD with the Conformer encoder has a much smaller number of parameters compared with the Transformer MD investigated by Dalmia~{\etal}\cite{dalmia-etal-2021-searchable} (67.2M vs. 135.3M) because we halved $\dmodel$.
Therefore, the Conformer-based Fast-MD models are more parameter-efficient.

\begin{table}[t]
    \centering
    \caption{Evaluation of searchability of better HI in Fast-MD on Fisher-CallHome Spanish corpus. $\beamst$ was set to 4 for all conditions. Transformer encoder was used.}
    \label{tab:asru2021_searchability}
    \vspace{-3mm}
    \begingroup
    \scalebox{0.88}{
    \begin{tabular}{lccccc} \toprule
    \multirow{3}{*}{Model} & \multicolumn{5}{c}{BLEU ($\uparrow$)} \\ \cmidrule{2-6}
    & \multicolumn{3}{c}{Fisher} & \multicolumn{2}{c}{CallHome} \\ \cmidrule(lr){2-4} \cmidrule(lr){5-6}
    & dev & dev2 & test & devtest & evltest \\
    \midrule
       MD + Fast-MD inference & 51.8 & 52.4 & 52.3 & 19.7 & 19.7 \\
       MD + Slow-MD inference & 53.8 & 54.5 & 54.3 & 21.5 & 21.5 \\
       \ + CTC scores & 54.4 & 54.8 & 54.3 & 21.8 & 21.8 \\
       \ \ ++ LM scores & \bf{55.1} & \bf{55.4} & \bf{54.7} & \bf{22.0} & \bf{21.9} \\
       \cmidrule(lr){1-6}

       Fast-MD (parallel HI) & 54.0 & 54.8 & 53.9 & 20.9 & 20.7 \\  %
       \ + Slow-MD inference & 54.4 & 54.7 & 54.5 & 21.2 & 21.7 \\
       \ \ ++ CTC scores & 54.6 & 55.4 & 54.5 & 21.5 & 21.8 \\
       \ \ \ +++ LM scores & \bf{54.9} & \bf{55.5} & \bf{54.8} & \bf{21.7} & \bf{21.8} \\
     \bottomrule
    \end{tabular}
    }
    \endgroup
\end{table}

\begin{figure}[t]
  \centering
  \includegraphics[width=0.99\linewidth]{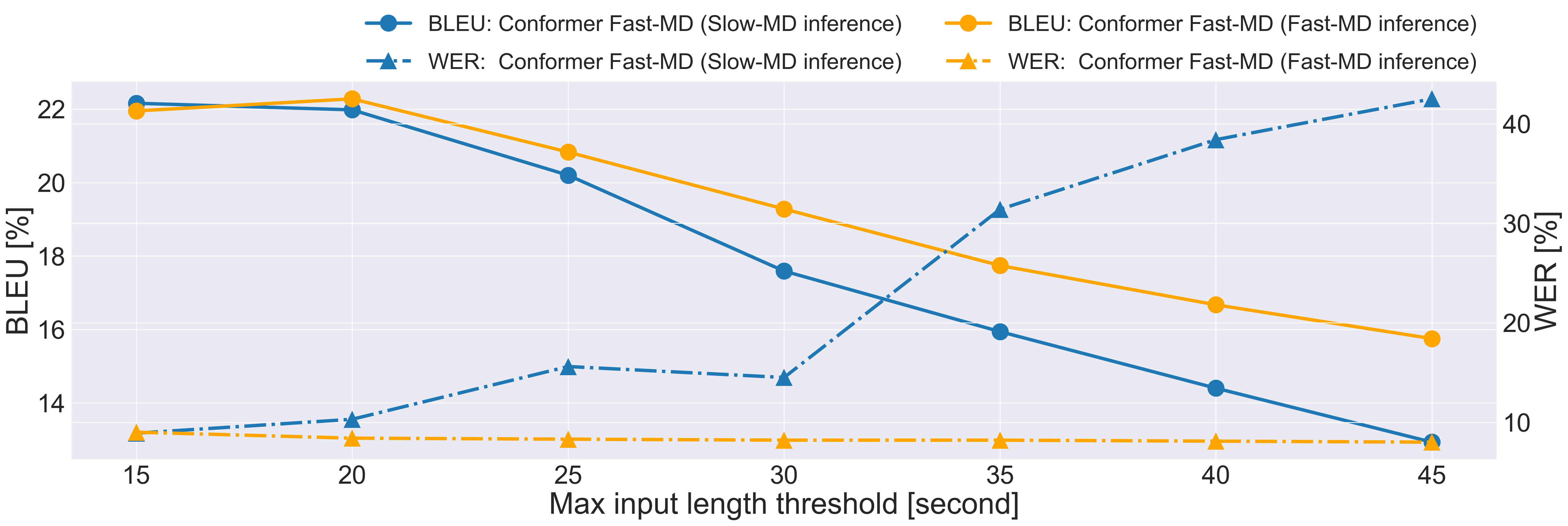}
  \caption{BLEU and WER as function of input-speech lengths on IWSLT \texttt{tst2019} set}
  \label{fig:asru2021_longform_fastmd_mustc}
  \vspace{-1mm}
\end{figure}

\vspace{-1mm}
\section{Analysis}
\subsection{Fast-MD does not lose searchability}\label{ssec:searchability_evaluation}
In Section~\ref{ssec:main_results}, we discussed how Fast-MD can recover a loss of translation quality caused by not carrying out beam search in the ASR sub-net.
We show that Fast-MD does not lose the searchability of better HI when the slow inference is accepted.
The results in Table~\ref{tab:asru2021_searchability} indicate that we can improve the BLEU scores of Fast-MD by integrating CTC and LM scores, which were comparative to those of the naïve MD model.

\vspace{-1mm}
\subsection{Fast-MD is robust against long-form speech}\label{ssec:fastmd_longform_evaluation}
It is well-known that ST results change significantly in accordance with audio segmentation~\cite{gaido2020contextualized,pham2020relative,potapczyk-przybysz-2020-srpols,gaido2021beyond,inaguma-etal-2021-espnet}.
Therefore, we evaluated the Conformer Fast-MD (parallel HI) trained on Must-C with the IWSLT \texttt{tst2019} set by changing the segmentation.
We simulated speech of various lengths up to 45 s on the basis of the segment-merging algorithm~\cite{inaguma-etal-2021-espnet}, where the entire speech in a session was split by a voice activity detection model~\cite{bredin2020pyannote} and multiple adjacent segments were concatenated until reaching the desired input length.
Figure~\ref{fig:asru2021_longform_fastmd_mustc} shows BLEU and WER of the same Fast-MD model while changing the decoding in the ASR sub-net.
We can see that WER with the Slow-MD inference without auxiliary CTC scores increased very quickly over 30 s and BLEU scores degraded accordingly.
On the other hand, WER with the Fast-MD inference did not degrade against long-form speech thanks to CTC, and the regression of BLEU scores was slower.
Therefore, we conclude that Fast-MD is robust against long-form speech.

\begin{table}[t]
    \centering
    \caption{Ablation study of Fast-MD on Fisher-\texttt{dev} set. Transformer encoder was used.}
    \label{tab:asru2021_ablation_study}
    \vspace{-3mm}
    \begingroup
    \scalebox{0.86}{
    \begin{tabular}{lc} \toprule
    Model & BLEU ($\uparrow$) \\
    \midrule
       MD + Fast-MD inference & 51.8 \\
       Fast-MD (parallel HI) w/ $\certhres=0.4$ & \bf{54.0} \\
       \ + ASR-PT (all ASR sub-net) & \bf{54.0} \\
       \ + $\certhres=0.2$ & 53.8\\
       \ + $\certhres=0.3$ & 53.7 \\
       \ + $\certhres=\infty$ (w/o CER thresholding) & 52.9 \\
       \ \ ++ ASR-PT (all ASR sub-net) & 53.4 \\
       \cmidrule(lr){1-2}

        Fast-MD (masked HI) w/ $\certhres=0.4$ & \bf{54.8} \\
        \ + w/o ASR-PT & 52.9 \\  %
        \ \ ++ train for more 50 epochs & 53.3 \\  %
     \bottomrule
    \end{tabular}
    }
    \endgroup
\end{table}

\vspace{-1mm}
\subsection{Ablation study}
Finally, we conducted an ablation study of Fast-MD on the Fisher-\texttt{dev} set shown in Table~\ref{tab:asru2021_ablation_study}.
We observed that $\certhres=0.4$ was best for the Transformer-based Fast-MD on this corpus, and $\certhres$ did not have a large impact on the BLEU score although both $\certhres=0.2$ and $\certhres=0.3$ improved it.
However, always sampling CTC outputs ($\certhres=\infty$) was not effective.
This is probably because the CER in the early training stage was high, leading to harmful training signals.
To reduce the CER in the early training stage, we initialized the ASR sub-net with a pre-trained ASR model (ASR-PT), but this was less effective than training from scratch with $\certhres=0.4$.
Therefore, we reason that CER thresholding is a type of curriculum learning~\cite{bengio2009curriculum} because CTC predictions are used gradually as training progresses.
However, ASR-PT was important for Fast-MD (masked HI).

\vspace{-1mm}
\section{Conclusion}\label{sec:conclusion}
We proposed Fast-MD to increase the decoding speed of the MD architecture for the E2E-ST task.
We generated HI from the ASR sub-net by NAR decoding led by CTC outputs.
We studied parallel HI and masked HI as NAR HI.
We also introduced sampling predictions from the CTC layer during training to reduce the gap in training and test conditions.
Experimental evaluations on three corpora showed that Fast-MD achieved comparable BLEU scores with 2$\times$ and 4$\times$ faster decoding than the naïve MD model on a GPU and CPU.
Integrating the Conformer encoder and InterCTC loss further boosted the translation quality without sacrificing decoding speed.\footnote{This work was partly supported by ASAPP and JHU HLTCOE.
This work used the Extreme Science and Engineering Discovery Environment (XSEDE)~\cite{towns2014xsede}, which is supported by NSF grant number ACI-1548562.
Specifically, it used the Bridges system~\cite{nystrom2015bridges}, supported by NSF grant ACI-1445606, at the Pittsburgh Supercomputing Center.
We also thank Jumon Nozaki for helpful discussions.}

\footnotesize
\bibliographystyle{IEEEbib}
\bibliography{reference}

\begin{thebibliography}{10}

\bibitem{ney1999speech}
Hermann Ney,
\newblock ``Speech translation: {C}oupling of recognition and translation,''
\newblock in {\em Proc. ICASSP}. IEEE, 1999, pp. 517--520.

\bibitem{weiss2017sequence}
Ron~J Weiss, Jan Chorowski, Navdeep Jaitly, Yonghui Wu, and Zhifeng Chen,
\newblock ``Sequence-to-sequence models can directly translate foreign
  speech,''
\newblock in {\em Proc. Interspeech}, 2017, pp. 2625--2629.

\bibitem{berard2018end}
Alexandre B{\'e}rard, Laurent Besacier, Ali~Can Kocabiyikoglu, and Olivier
  Pietquin,
\newblock ``End-to-end automatic speech translation of audiobooks,''
\newblock in {\em Proc. ICASSP}. IEEE, 2018, pp. 6224--6228.

\bibitem{bansal-etal-2019-pre}
Sameer Bansal, Herman Kamper, Karen Livescu, Adam Lopez, and Sharon Goldwater,
\newblock ``Pre-training on high-resource speech recognition improves
  low-resource speech-to-text translation,''
\newblock in {\em Proc. NAACL-HLT}, 2019, pp. 58--68.

\bibitem{bahar2019comparative}
Parnia Bahar, Tobias Bieschke, and Hermann Ney,
\newblock ``A comparative study on end-to-end speech to text translation,''
\newblock in {\em Proc. ASRU}. IEEE, 2019, pp. 792--799.

\bibitem{liu2020synchronous}
Yuchen Liu, Jiajun Zhang, Hao Xiong, Long Zhou, Zhongjun He, Hua Wu, Haifeng
  Wang, and Chengqing Zong,
\newblock ``Synchronous speech recognition and speech-to-text translation with
  interactive decoding,''
\newblock in {\em Proc. AAAI}, 2020, pp. 8417--8424.

\bibitem{le-etal-2020-dual}
Hang Le, Juan Pino, Changhan Wang, Jiatao Gu, Didier Schwab, and Laurent
  Besacier,
\newblock ``Dual-decoder transformer for joint automatic speech recognition and
  multilingual speech translation,''
\newblock in {\em Proc. COLING}, 2020, pp. 3520--3533.

\bibitem{dong2021consecutive}
Qianqian Dong, Mingxuan Wang, Hao Zhou, Shuang Xu, Bo~Xu, and Lei Li,
\newblock ``Consecutive decoding for speech-to-text translation,''
\newblock in {\em Proc. AAAI}, 2021.

\bibitem{anastasopoulos-chiang-2018-tied}
Antonios Anastasopoulos and David Chiang,
\newblock ``Tied multitask learning for neural speech translation,''
\newblock in {\em Proc. NAACL-HLT}, 2018, pp. 82--91.

\bibitem{sperber-etal-2019-attention}
Matthias Sperber, Graham Neubig, Jan Niehues, and Alex Waibel,
\newblock ``Attention-passing models for robust and data-efficient end-to-end
  speech translation,''
\newblock {\em Transactions of the Association for Computational Linguistics},
  vol. 7, 2019.

\bibitem{sung2019towards}
Tzu-Wei Sung, Jun-You Liu, Hung-yi Lee, and Lin-shan Lee,
\newblock ``Towards end-to-end speech-to-text translation with two-pass
  decoding,''
\newblock in {\em Proc. ICASSP}. IEEE, 2019, pp. 7175--7179.

\bibitem{dalmia-etal-2021-searchable}
Siddharth Dalmia, Brian Yan, Vikas Raunak, Florian Metze, and Shinji Watanabe,
\newblock ``Searchable hidden intermediates for end-to-end models of
  decomposable sequence tasks,''
\newblock in {\em Proc. NAACL-HLT}, 2021, pp. 1882--1896.

\bibitem{dalmia2019enforcing}
Siddharth Dalmia, Abdelrahman Mohamed, Mike Lewis, Florian Metze, and Luke
  Zettlemoyer,
\newblock ``Enforcing encoder-decoder modularity in sequence-to-sequence
  models,''
\newblock {\em arXiv preprint arXiv:1911.03782}, 2019.

\bibitem{shi-etal-2021-highland}
Jiatong Shi, Jonathan~D. Amith, Xuankai Chang, Siddharth Dalmia, Brian Yan, and
  Shinji Watanabe,
\newblock ``{H}ighland {P}uebla {N}ahuatl speech translation corpus for
  endangered language documentation,''
\newblock in {\em Proc. AmericasNLP}, 2021, pp. 53--63.

\bibitem{ctc_graves}
Alex Graves, Santiago Fern{\'a}ndez, Faustino Gomez, and J{\"u}rgen
  Schmidhuber,
\newblock ``Connectionist temporal classification: {L}abelling unsegmented
  sequence data with recurrent neural networks,''
\newblock in {\em Proc. ICML}, 2006, pp. 369--376.

\bibitem{liu2020bridging}
Yuchen Liu, Junnan Zhu, Jiajun Zhang, and Chengqing Zong,
\newblock ``Bridging the modality gap for speech-to-text translation,''
\newblock {\em arXiv preprint arXiv:2010.14920}, 2020.

\bibitem{gaido-etal-2021-ctc}
Marco Gaido, Mauro Cettolo, Matteo Negri, and Marco Turchi,
\newblock ``{CTC}-based compression for direct speech translation,''
\newblock in {\em Proc. EACL}, 2021, pp. 690--696.

\bibitem{zeng-etal-2021-realtrans}
Xingshan Zeng, Liangyou Li, and Qun Liu,
\newblock ``{R}eal{T}ran{S}: {E}nd-to-end simultaneous speech translation with
  convolutional weighted-shrinking {Transformer},''
\newblock in {\em Findings of the Association for Computational Linguistics:
  ACL-IJCNLP 2021}, 2021, pp. 2461--2474.

\bibitem{zhang-etal-2020-adaptive}
Biao Zhang, Ivan Titov, Barry Haddow, and Rico Sennrich,
\newblock ``Adaptive feature selection for end-to-end speech translation,''
\newblock in {\em Findings of the Association for Computational Linguistics:
  EMNLP 2020}, 2020, pp. 2533--2544.

\bibitem{xu-etal-2021-stacked}
Chen Xu, Bojie Hu, Yanyang Li, Yuhao Zhang, Shen Huang, Qi~Ju, Tong Xiao, and
  Jingbo Zhu,
\newblock ``Stacked acoustic-and-textual encoding: {I}ntegrating the
  pre-trained models into speech translation encoders,''
\newblock in {\em Proc. ACL}, 2021, pp. 2619--2630.

\bibitem{inaguma2021orthros}
Hirofumi Inaguma, Yosuke Higuchi, Kevin Duh, Tatsuya Kawahara, and Shinji
  Watanabe,
\newblock ``Orthros: {N}on-autoregressive end-to-end speech translation with
  dual-decoder,''
\newblock in {\em Proc. ICASSP}. IEEE, 2021, pp. 7503--7507.

\bibitem{chuang-etal-2021-investigating}
Shun-Po Chuang, Yung-Sung Chuang, Chih-Chiang Chang, and Hung-yi Lee,
\newblock ``Investigating the reordering capability in {CTC}-based
  non-autoregressive end-to-end speech translation,''
\newblock in {\em Findings of the Association for Computational Linguistics:
  ACL-IJCNLP 2021}, 2021, pp. 1068--1077.

\bibitem{higuchi2020mask}
Yosuke Higuchi, Shinji Watanabe, Nanxin Chen, Tetsuji Ogawa, and Tetsunori
  Kobayashi,
\newblock ``Mask {CTC}: {N}on-autoregressive end-to-end {ASR} with {CTC} and
  mask predict,''
\newblock in {\em Proc. Interspeech}, 2020, pp. 3655--3659.

\bibitem{ghazvininejad-etal-2019-mask}
Marjan Ghazvininejad, Omer Levy, Yinhan Liu, and Luke Zettlemoyer,
\newblock ``Mask-predict: {P}arallel decoding of conditional masked language
  models,''
\newblock in {\em Proc. EMNLP}, 2019, pp. 6114--6123.

\bibitem{williams1989learning}
Ronald~J Williams and David Zipser,
\newblock ``A learning algorithm for continually running fully recurrent neural
  networks,''
\newblock {\em Neural computation}, vol. 1, no. 2, pp. 270--280, 1989.

\bibitem{gulati2020}
Anmol Gulati, James Qin, Chung-Cheng Chiu, Niki Parmar, Yu~Zhang, Jiahui Yu,
  Wei Han, Shibo Wang, Zhengdong Zhang, Yonghui Wu, and Ruoming Pang,
\newblock ``Conformer: {C}onvolution-augmented {Transformer} for speech
  recognition,''
\newblock in {\em Proc. Interspeech}, 2020, pp. 5036--5040.

\bibitem{lee2021intermediate}
Jaesong Lee and Shinji Watanabe,
\newblock ``Intermediate loss regularization for {CTC}-based speech
  recognition,''
\newblock in {\em Proc. ICASSP}. IEEE, 2021, pp. 6224--6228.

\bibitem{karita2019comparative}
Shigeki Karita, Nanxin Chen, Tomoki Hayashi, Takaaki Hori, Hirofumi Inaguma,
  Ziyan Jiang, Masao Someki, Nelson Enrique~Yalta Soplin, Ryuichi Yamamoto,
  Xiaofei Wang, et~al.,
\newblock ``A comparative study on {Transformer} vs {RNN} in speech
  applications,''
\newblock in {\em Proc. ASRU}. IEEE, 2019, pp. 499--456.

\bibitem{hybrid_ctc_attention}
Shinji Watanabe, Takaaki Hori, Suyoun Kim, John~R Hershey, and Tomoki Hayashi,
\newblock ``Hybrid {CTC}/attention architecture for end-to-end speech
  recognition,''
\newblock {\em IEEE Journal of Selected Topics in Signal Processing}, vol. 11,
  no. 8, pp. 1240--1253, 2017.

\bibitem{song2021non}
Xingchen Song, Zhiyong Wu, Yiheng Huang, Chao Weng, Dan Su, and Helen Meng,
\newblock ``Non-autoregressive transformer {ASR} with {CTC}-enhanced decoder
  input,''
\newblock in {\em Proc. ICASSP}. IEEE, 2021, pp. 5894--5898.

\bibitem{fan2021cass}
Ruchao Fan, Wei Chu, Peng Chang, and Jing Xiao,
\newblock ``{CASS-NAT}: {CTC} alignment-based single step non-autoregressive
  transformer for speech recognition,''
\newblock in {\em Proc. ICASSP}. IEEE, 2021, pp. 5889--5893.

\bibitem{tian2020one}
Zhengkun Tian, Jiangyan Yi, Ye~Bai, Jianhua Tao, Shuai Zhang, and Zhengqi Wen,
\newblock ``One in a hundred: {S}elect the best predicted sequence from
  numerous candidates for streaming speech recognition,''
\newblock {\em arXiv preprint arXiv:2010.14791}, 2020.

\bibitem{wang2021wnars}
Zhichao Wang, Wenwen Yang, Pan Zhou, and Wei Chen,
\newblock ``{WNARS}: {WFST} based non-autoregressive streaming end-to-end
  speech recognition,''
\newblock {\em arXiv preprint arXiv:2104.03587}, 2021.

\bibitem{zhang2020unified}
Binbin Zhang, Di~Wu, Zhuoyuan Yao, Xiong Wang, Fan Yu, Chao Yang, Liyong Guo,
  Yaguang Hu, Lei Xie, and Xin Lei,
\newblock ``Unified streaming and non-streaming two-pass end-to-end model for
  speech recognition,''
\newblock {\em arXiv preprint arXiv:2012.05481}, 2020.

\bibitem{higuchi2021improved}
Yosuke Higuchi, Hirofumi Inaguma, Shinji Watanabe, Tetsuji Ogawa, and Tetsunori
  Kobayashi,
\newblock ``Improved {Mask-CTC} for non-autoregressive end-to-end {ASR},''
\newblock in {\em Proc. ICASSP}. IEEE, 2021, pp. 8363--8367.

\bibitem{tjandra2020deja}
Andros Tjandra, Chunxi Liu, Frank Zhang, Xiaohui Zhang, Yongqiang Wang, Gabriel
  Synnaeve, Satoshi Nakamura, and Geoffrey Zweig,
\newblock ``Deja-vu: {D}ouble feature presentation and iterated loss in deep
  transformer networks,''
\newblock in {\em Proc. ICASSP}. IEEE, 2020, pp. 6899--6903.

\bibitem{nozaki2021relaxing}
Jumon Nozaki and Tatsuya Komatsu,
\newblock ``Relaxing the conditional independence assumption of {CTC}-based
  {ASR} by conditioning on intermediate predictions,''
\newblock in {\em Proc. Interspeech}, 2021, pp. 3735--3739.

\bibitem{ng2021pushing}
Edwin~G Ng, Chung-Cheng Chiu, Yu~Zhang, and William Chan,
\newblock ``Pushing the limits of non-autoregressive speech recognition,''
\newblock in {\em Proc. Interspeech}, 2021, pp. 3725--3729.

\bibitem{inaguma-etal-2020-espnet}
Hirofumi Inaguma, Shun Kiyono, Kevin Duh, Shigeki Karita, Nelson Yalta, Tomoki
  Hayashi, and Shinji Watanabe,
\newblock ``{ESP}net-{ST}: All-in-one speech translation toolkit,''
\newblock in {\em Proc. ACL: System Demonstrations}, 2020, pp. 302--311.

\bibitem{fisher_callhome}
Matt Post, Gaurav Kumar, Adam Lopez, Damianos Karakos, Chris Callison-Burch,
  and Sanjeev Khudanpur,
\newblock ``Improved speech-to-text translation with the {Fisher} and {Callhome
  Spanish--English} speech translation corpus,''
\newblock in {\em Proc. IWSLT}, 2013.

\bibitem{papineni-etal-2002-bleu}
Kishore Papineni, Salim Roukos, Todd Ward, and Wei-Jing Zhu,
\newblock ``{B}leu: a method for automatic evaluation of machine translation,''
\newblock in {\em Proc. ACL}. 2002, pp. 311--318, Association for Computational
  Linguistics.

\bibitem{libri_trans}
Ali~Can Kocabiyikoglu, Laurent Besacier, and Olivier Kraif,
\newblock ``Augmenting {Librispeech} with {French} translations: {A} multimodal
  corpus for direct speech translation evaluation,''
\newblock in {\em Proc. LREC}, 2018.

\bibitem{mustc}
Mattia~A. Di~Gangi, Roldano Cattoni, Luisa Bentivogli, Matteo Negri, and Marco
  Turchi,
\newblock ``{M}u{ST}-{C}: a {M}ultilingual {S}peech {T}ranslation {C}orpus,''
\newblock in {\em Proc. NAACL-HLT}, 2019, pp. 2012--2017.

\bibitem{kaldi}
Daniel Povey, Arnab Ghoshal, Gilles Boulianne, Lukas Burget, Ondrej Glembek,
  Nagendra Goel, Mirko Hannemann, Petr Motlicek, Yanmin Qian, Petr Schwarz,
  et~al.,
\newblock ``The kaldi speech recognition toolkit,''
\newblock in {\em Proc. ASRU}. IEEE, 2011.

\bibitem{speed_perturbation}
Tom Ko, Vijayaditya Peddinti, Daniel Povey, and Sanjeev Khudanpur,
\newblock ``Audio augmentation for speech recognition,''
\newblock in {\em Proc. Interspeech}, 2015, pp. 3586--3589.

\bibitem{specaugment}
Daniel~S Park, William Chan, Yu~Zhang, Chung-Cheng Chiu, Barret Zoph, Ekin~D
  Cubuk, and Quoc~V Le,
\newblock ``{SpecAugment}: {A} simple data augmentation method for automatic
  speech recognition,''
\newblock in {\em Proc. Interspeech}, 2019, pp. 2613--2617.

\bibitem{koehn-etal-2007-moses}
Philipp Koehn, Hieu Hoang, Alexandra Birch, Chris Callison-Burch, Marcello
  Federico, Nicola Bertoldi, Brooke Cowan, Wade Shen, Christine Moran, Richard
  Zens, Chris Dyer, Ond{\v{r}}ej Bojar, Alexandra Constantin, and Evan Herbst,
\newblock ``{M}oses: {O}pen source toolkit for statistical machine
  translation,''
\newblock in {\em Proc. ACL: Demo and Poster Sessions}, 2007, pp. 177--180.

\bibitem{sennrich-etal-2016-neural}
Rico Sennrich, Barry Haddow, and Alexandra Birch,
\newblock ``Neural machine translation of rare words with subword units,''
\newblock in {\em Proc. ACL}, 2016, pp. 1715--1725.

\bibitem{kudo-richardson-2018-sentencepiece}
Taku Kudo and John Richardson,
\newblock ``{S}entence{P}iece: A simple and language independent subword
  tokenizer and detokenizer for neural text processing,''
\newblock in {\em Proc. EMNLP: System Demonstrations}, 2018, pp. 66--71.

\bibitem{vaswani2017attention}
Ashish Vaswani, Noam Shazeer, Niki Parmar, Jakob Uszkoreit, Llion Jones,
  Aidan~N Gomez, {\L}ukasz Kaiser, and Illia Polosukhin,
\newblock ``Attention is all you need,''
\newblock in {\em Proc. NIPS}, 2017, pp. 5998--6008.

\bibitem{guo2021recent}
Pengcheng Guo, Florian Boyer, Xuankai Chang, Tomoki Hayashi, Yosuke Higuchi,
  Hirofumi Inaguma, Naoyuki Kamo, Chenda Li, Daniel Garcia-Romero, Jiatong Shi,
  et~al.,
\newblock ``Recent developments on {ESPnet} toolkit boosted by {Conformer},''
\newblock in {\em Proc. ICASSP}. IEEE, 2021, pp. 5874--5878.

\bibitem{adam}
Diederik Kingma and Jimmy Ba,
\newblock ``Adam: {A} method for stochastic optimization,''
\newblock {\em Proc. ICLR}, 2015.

\bibitem{seki2019}
Hiroshi Seki, Takaaki Hori, Shinji Watanabe, Niko Moritz, and Jonathan~Le Roux,
\newblock ``Vectorized beam search for {CTC}-attention-based speech
  recognition,''
\newblock in {\em Proc. Interspeech}, 2019, pp. 3825--3829.

\bibitem{post-2018-call}
Matt Post,
\newblock ``A call for clarity in reporting {BLEU} scores,''
\newblock in {\em Proc. WMT: Research Papers}, 2018, pp. 186--191.

\bibitem{liu2019end}
Yuchen Liu, Hao Xiong, Zhongjun He, Jiajun Zhang, Hua Wu, Haifeng Wang, and
  Chengqing Zong,
\newblock ``End-to-end speech translation with knowledge distillation,''
\newblock in {\em Proc. Interspeech}, 2019, pp. 1128--1132.

\bibitem{wang2020bridging}
Chengyi Wang, Yu~Wu, Shujie Liu, Zhenglu Yang, and Ming Zhou,
\newblock ``Bridging the gap between pre-training and fine-tuning for
  end-to-end speech translation,''
\newblock in {\em Proc. AAAI}, 2020, pp. 9161--9168.

\bibitem{zhao-etal-2021-neurst}
Chengqi Zhao, Mingxuan Wang, Qianqian Dong, Rong Ye, and Lei Li,
\newblock ``{N}eur{ST}: {N}eural speech translation toolkit,''
\newblock in {\em Proc. ACL: System Demonstrations}, 2021, pp. 55--62.

\bibitem{wang-etal-2020-curriculum}
Chengyi Wang, Yu~Wu, Shujie Liu, Ming Zhou, and Zhenglu Yang,
\newblock ``Curriculum pre-training for end-to-end speech translation,''
\newblock in {\em Proceedings ACL}, 2020, pp. 3728--3738.

\bibitem{dong2021listen}
Qianqian Dong, Rong Ye, Mingxuan Wang, Hao Zhou, Shuang Xu, Bo~Xu, and Lei Li,
\newblock ``{“Listen, Understand and Translate”}: {T}riple supervision
  decouples end-to-end speech-to-text translation,''
\newblock in {\em Proc. AAAI}, 2021.

\bibitem{wang-etal-2020-fairseq}
Changhan Wang, Yun Tang, Xutai Ma, Anne Wu, Dmytro Okhonko, and Juan Pino,
\newblock ``Fairseq {S}2{T}: Fast speech-to-text modeling with fairseq,''
\newblock in {\em Proc. AACL: System Demonstrations}, 2020, pp. 33--39.

\bibitem{pham2020relative}
Ngoc-Quan Pham, Thanh-Le Ha, Tuan-Nam Nguyen, Thai-Son Nguyen, Elizabeth
  Salesky, Sebastian Stüker, Jan Niehues, and Alex Waibel,
\newblock ``Relative positional encoding for speech recognition and direct
  translation,''
\newblock in {\em Proc. Interspeech}, 2020, pp. 31--35.

\bibitem{inaguma-etal-2021-source}
Hirofumi Inaguma, Tatsuya Kawahara, and Shinji Watanabe,
\newblock ``Source and target bidirectional knowledge distillation for
  end-to-end speech translation,''
\newblock in {\em Proc. NAACL-HLT}, 2021, pp. 1872--1881.

\bibitem{gaido2020contextualized}
Marco Gaido, Mattia A.~Di Gangi, Matteo Negri, Mauro Cettolo, and Marco Turchi,
\newblock ``Contextualized translation of automatically segmented speech,''
\newblock in {\em Proc. Interspeech}, 2020, pp. 1471--1475.

\bibitem{potapczyk-przybysz-2020-srpols}
Tomasz Potapczyk and Pawel Przybysz,
\newblock ``{SRPOL}{'}s system for the {IWSLT} 2020 end-to-end speech
  translation task,''
\newblock in {\em Proc. IWSLT}, 2020, pp. 89--94.

\bibitem{gaido2021beyond}
Marco Gaido, Matteo Negri, Mauro Cettolo, and Marco Turchi,
\newblock ``Beyond voice activity detection: {H}ybrid audio segmentation for
  direct speech translation,''
\newblock {\em arXiv preprint arXiv:2104.11710}, 2021.

\bibitem{inaguma-etal-2021-espnet}
Hirofumi Inaguma, Brian Yan, Siddharth Dalmia, Pengcheng Guo, Jiatong Shi,
  Kevin Duh, and Shinji Watanabe,
\newblock ``{ESP}net-{ST} {IWSLT} 2021 offline speech translation system,''
\newblock in {\em Proc. IWSLT}, 2021, pp. 100--109.

\bibitem{bredin2020pyannote}
Herv{\'e} Bredin, Ruiqing Yin, Juan~Manuel Coria, Gregory Gelly, Pavel
  Korshunov, Marvin Lavechin, Diego Fustes, Hadrien Titeux, Wassim Bouaziz, and
  Marie-Philippe Gill,
\newblock ``Pyannote.audio: neural building blocks for speaker diarization,''
\newblock in {\em Proc. ICASSP}, 2020, pp. 7124--7128.

\bibitem{bengio2009curriculum}
Yoshua Bengio, J{\'e}r{\^o}me Louradour, Ronan Collobert, and Jason Weston,
\newblock ``Curriculum learning,''
\newblock in {\em Proc. ICML}, 2009, pp. 41--48.

\bibitem{towns2014xsede}
John Towns, Timothy Cockerill, Maytal Dahan, Ian Foster, Kelly Gaither, et~al.,
\newblock ``Xsede: Accelerating scientific discovery,''
\newblock {\em Computing in Science Engineering}, vol. 16, no. 5, pp. 62--74,
  2014.

\bibitem{nystrom2015bridges}
Nicholas~A Nystrom, Michael~J Levine, Ralph~Z Roskies, and J~Ray Scott,
\newblock ``Bridges: a uniquely flexible {HPC} resource for new communities and
  data analytics,''
\newblock in {\em Proc. 2015 XSEDE Conference: Scientific Advancements Enabled
  by Enhanced Cyberinfrastructure}, 2015.

\end{thebibliography}

\end{document}